\documentclass[aps,superscriptaddress,amsmath,amssymb,floatfix,showpacs,notitlepage,nofootinbib,preprintnumbers]{revtex4-1}
\usepackage{graphicx,graphics,setspace,epsfig,color}
\usepackage[letterpaper, dvips,width=7.5in,height=8.5in,includemp=false]{geometry}
\usepackage[vcentermath]{}
\usepackage{epsf}
\usepackage{amscd}
\usepackage{amsmath}
\usepackage{graphicx}% Include figure files
\usepackage{dcolumn}% Align table columns on decimal point
\usepackage{bm}% bold math
\usepackage{amssymb}
\usepackage{epsfig}
\usepackage{slashed}
\usepackage{lmodern} 
\usepackage[T1]{fontenc}

\newcommand{\beq}{\begin{equation}}
\newcommand{\eeq}{\end{equation}}
\newcommand{\bea}{\begin{eqnarray}}
\newcommand{\eea}{\end{eqnarray}}

\newcommand{\prince} {\mu_5}

\newcommand\eqn[1]{\label{eq:#1}} 
\newcommand\eq[1]{eq. (\ref{eq:#1})} 
\newcommand\Eq[1]{Eq.~(\ref{eq:#1})} 
\newcommand{\bfk}{{\mathbf k}}

\newcommand{\eV}{{\rm ~eV }}

\newcommand{\MeV}{{\rm ~MeV }}
\newcommand{\curl}{{\nabla \times}}

\newcommand{\CM}{{\cal M}}

\newcommand{\bp}{{\mathbf{p}}}

\newcommand{\bk}{{\mathbf{k}}}

\newcommand\ket[1]{| #1 \rangle}

\newcommand\expect[3]{\langle #1|#2|#3\rangle}
\begin{document}

\preprint{INT-PUB-14-039}

\title[title]{The Role of the Electron Mass in Damping  Chiral Magnetic Instability in Supernova and Neutron Stars}
\author{Dorota Grabowska}
 \email{grabow@uw.edu}
\affiliation{Institute for Nuclear Theory, University of Washington, Seattle, WA}
\author{David B. Kaplan}
 \email{dbkaplan@uw.edu}
\affiliation{Institute for Nuclear Theory, University of Washington, Seattle, WA}
\author{ Sanjay Reddy}
 \email{sareddy@uw.edu}
\affiliation{Institute for Nuclear Theory, University of Washington, Seattle, WA}

%\date{}%
%\dedicatory{}%
%\commby{}%
% ----------------------------------------------------------------

%\cite{Ohnishi:2014uea}
\begin{abstract}
We show that the nonzero electron mass plays a critical role in determining the magnetic properties of neutron stars, making it impossible  to generate the chiral charge density needed to trigger a strong chiral magnetic instability during the core collapse of supernovae. This instability has been proposed as a plausible mechanism for generating extremely large helical magnetic fields in neutron stars  at their birth; the mechanism relies  on the generation of a large non-equilibrium chiral charge density via electron capture reactions that selectively deplete left-handed electrons during core-collapse and the early evolution of the protoneutron star. Our calculation shows that the electron chirality violation rate induced by Rutherford scattering, despite being suppressed by the smallness of the electron mass relative to the electron chemical potential,  is still fast compared to the weak interaction electron capture rate. The resulting asymmetry between right and left-handed electron densities is therefore never able to attain an astrophysically relevant magnitude.             
\end{abstract}

\maketitle
The inference of extreme surface magnetic fields $B_S\simeq 10^{14}-10^{15}$ G  from observations of a class of neutron stars called magnetars \cite{McGillCatalog:2014} raises many questions about how and when such fields are generated. In the conventional scenario, they are expected to arise either due to strong hydrodynamical or magnetohydrodynamic instabilities during core-collapse supernova, or during the early evolution of the proto-neutron star \cite{Duncan:1992hi,Akiyama:2003,Endeve:2012}. Other mechanisms, which rely on a spontaneous magnetization of the ground state of strongly interacting matter at extreme density, have also been proposed; these remain speculative due to large theoretical uncertainties. Recently in \cite{Ohnishi:2014uea}, it was suggested that chiral magnetic instability (CMI) \cite{Vilenkin:1980fu} could be used to generate large fields. In this intriguing scenario, a net chiral charge is produced during core collapse of the progenitor star. As matter is compressed during core collapse, left-handed electrons are captured by protons due to the weak interaction, which results in an imbalance between the Fermi energies of left-handed and right-handed electrons.  This imbalance triggers an instability that equilibrates the two chiralities and the released energy drives the growth of a coherent magnetic field.  Key to this analysis is the assumption that the electron mass, which explicitly violates chirality, can be neglected \cite{Ohnishi:2014uea}. The authors argue that this is a reasonable approximation because the electron mass $m_e=0.51$ MeV is much smaller than the typical electron Fermi momentum $p_{\rm Fe}\simeq 100 $ MeV encountered in supernova and neutron stars. We revisit this assumption here and claim that in fact the electron mass cannot be neglected as it leads to chiral charge equilibration much faster than the weak interactions can create an asymmetry, and that therefore this mechanism does not lead to astrophysical interesting magnetic fields.

We start by reviewing the chiral magnetic instability for massless electrons with only electromagnetic interactions. In this case, chiral symmetry is only violated by quantum effects (the anomaly), and at the classical level left and right handed electron numbers are separately conserved. Asbsent the chiral anomaly, inverse beta decay during core collapse of the neutron star progenitor leads to a net chiral charge in the resultant neutron star.
Already in 1980, Vilenkin \cite{Vilenkin:1980fu} realized that a net chiral charge density in the plasma can trigger an instability, now called the chiral magnetic instability (CMI),  by inducing a contribution to the  electric current  proportional to the magnetic field
\beq 
\vec{J}=\frac{2 \alpha}{\pi} \prince \vec{B}    \,, 
\eqn{j_B} 
\eeq
where $\prince=\mu_R-\mu_L$ is the chemical potential associated with the chiral charge density, and $\mu_R$ and $\mu_L$ are the chemical potentials associated with the right and left handed massless particles and $\alpha=e^2/4\pi$ is the fine structure constant. We will refer to this as the chiral magnetic current; it can be derived from a parity violating effective action for the gauge fields (in the plasma rest frame) of the form 
$$
\int\, d^4x\, d^4y\,  g(x-y) \epsilon^{0ijk} A_i(y)\partial_j A_k(x)$$ 
where $g(x-y)$ is in general nonlocal and proportional to $\mu_5$ with the chiral magnetic current arising from the leading term in a derivative expansion of $g$.

The origin of the chiral magnetic current is easy to understand:  in a constant magnetic field electrons occupy Landau levels, where each Landau level can be viewed as a 1+1 dimensional Dirac fermion traveling along the direction of the magnetic field; the excited levels contain electrons of both spin polarizations, while the lowest Landau level only contains electrons with spin anti-aligned with the field. At nonzero $\mu_5$ it follows that there is a difference between the density of particles  in the lowest Landau level moving parallel to the magnetic field (LH chirality) versus antiparallel (RH chirality), and hence there exists an  electric current in the direction of the magnetic field, $\vec{B}$. It is given by the $1+1$ dimensional current density in the magnetic field direction, $ (e\mu_5/2\pi) $, times the transverse density of the lowest Landau orbits, $(eB/\pi)$ (see derivation in \cite{Kaplan:2009yg}, for example).   Nonzero $\mu_5$ also forces a chiral asymmetry  in the excited Landau levels, but as these levels contain electrons of both polarizations they do not contribute to the electric current.    No mention has been made of the anomaly, but the Landau level picture of the anomaly in 3+1 dimensions shows that the two are intimately related \footnote{See  J. Preskill's lecture notes on Quantum Chromodynamics at http://www.theory.caltech.edu/\textasciitilde preskill/notes.html, pp. 3.43-3.45, or  else the derivation in \cite{Kaplan:2009yg}.}. 

When modified to incorporate the chiral magnetic current, \Eq{j_B}, and  finite electrical conductivity, Maxwell's equations read
\bea 
\frac{\partial \vec{B}}{\partial t} &=& -\curl \vec{E} \\
\curl \vec{B} -\frac{\partial \vec{E}}{\partial t}&=& \sigma~\vec{E} + \frac{2\alpha}{\pi}\mu_5~\vec{B} \equiv \vec J\,,
\label{eq:Maxwell_1}
\eea   
where $\sigma$ is the electrical conductivity, scaling as $\mu_e/\alpha$. Assuming constant $\sigma$ and $\mu_5$, and ignoring the $ \partial \vec{E}/\partial t$ term (justified below)  we combine the above equations to obtain
\beq
\frac{\partial \vec{B}}{\partial t} =\frac{1}{\sigma}~\nabla^2\vec{B} + \frac{2\alpha}{\pi \sigma}~\mu_5 \curl\vec{B} \,, 
\label{eq:cme}
\eeq
which describes the time evolution of $ \vec{B}$ in the presence of the chiral magnetic current. 
The unstable modes are characterized by the vector potential 
\beq
  \vec{A}_\pm = (\hat{x} \pm i\hat{y})~e^{(i k z -i \omega t)}\ ,
\eqn{helA}
  \eeq
which  corresponds to electric fields $\vec E_\pm = i \omega   \vec{A}_\pm$ and magnetic fields $\vec B_\pm = \pm k   \vec{A}_\pm$, where the $\pm$ subscript denotes the helicity of the fields for positive $k$. The wavenumber $k$ and the frequency $\Re[\omega]$ are constants.  \Eq{Maxwell_1} has exponentially growing solutions, whose helicity depend on the sign of $\mu_5$,  with amplitude 
\beq 
B_k(t) = B_k(0)e^{t k(2k_\star -k)/\sigma} \ ,\qquad k_\star  = \frac{\alpha\mu_5}{\pi}   
\label{eq:boft}
\eeq
for $0<k<2k_\star $,
where $B_k(0)$ is the initial magnetic field --  either a thermal fluctuation, or the field inherited from the progenitor star.  Note that the terms kept in \Eq{cme} are proportional to $k_\star^2/\sigma$, while the term $\partial\vec E/\partial t$ neglected in going from \Eq{Maxwell_1} to \Eq{cme} is smaller by a factor of  $\omega/k = k_\star/\sigma = O(\alpha^2)$;  similarly the neglected plasma frequency of the photon has a negligible effect on the growing mode solution.
The maximally  unstable mode  occurs for  $k =k_\star $, with that mode growing as 
\beq
B_{\star}(t) = B_{\star}(0)\exp{(\Gamma_{\rm CMI}\,t)}\ ,\qquad \Gamma_{\rm CMI} = \frac{k_\star ^2}{\sigma} =\frac{\alpha^2 \mu_5^2}{\pi^2\sigma}\ .
\label{eq:cmi}
\eeq
 For a recent discussion of this instability in the context of the high temperature plasma encountered in the early universe see Ref.~\cite{Boyarsky:2012,Tashiro:2012mf}.

The local evolution of the chiral charge density is described by the anomaly equation 
\beq
\partial_\mu j^\mu_5= -\frac{\alpha}{2\pi} F_{\alpha\beta} \tilde F^{\alpha\beta} = -\partial_\mu K^\mu \ , \qquad \qquad K^\mu = \frac{\alpha}{\pi}\epsilon^{\mu\alpha\beta\gamma}A_\alpha\partial_\beta A_\gamma \ .
\eqn{anomaly} 
\eeq
Integrating over space and assuming fields vanish at spatial infinity yields the conservation law
\beq
\frac{d\,}{dt}\left( n_5+\frac{\alpha}{\pi}H\right) = 0\ ,\qquad H =\frac{1}{V}\int d^3x \,\vec A\cdot\vec B\ ,
\eeq
where $n_5 = N_5/V$ is the average chiral charge density,  $V$ is the volume,  and $H$  is the gauge invariant ``helicity density".  Note that a time-dependent helicity implies a nonzero electric field, and thus the above equation can be simply understood as the conventional effect of an electric field changing the momenta of electrons in the lowest Landau level. 

Since the field \Eq{helA} has nonzero helicity,   the growth of the unstable mode converts electron chiral charge density $n_5$ into electromagnetic helicity $H$  at a rate 
\beq 
\frac{\partial n_5}{\partial t}=-\frac{\alpha}{ \pi}\frac{d H}{d t} =- \frac{2\alpha \Gamma_{\rm CMI}}{\pi k_\star }~B_\star(t)^2 =- \frac{2\alpha^2\mu_5}{\pi^2\sigma}~B_\star(t)^2\equiv -\Gamma_B n_5 \,,
\label{eq:hdot}
\eeq
where $B_\star(t)$ is given in \eq{cmi}.
The free energy in the magnetic field is supplied by the imbalance of Fermi energy between left and right handed electrons. In time, $\mu_5$ is driven to zero locally, and a global helical magnetic field that spontaneously breaks rotational symmetry  is generated. As we elaborate on below, this is the phenomena essential to the proposed mechanism for generating large magnetic fields during the supernova in Ref.~\cite{Ohnishi:2014uea}.  However, one immediately sees a problem with using the CMI to directly generate large coherent magnetic fields on astrophysical scales:  for long wavelength magnetic fields, $k_\star$ must be exceedingly small compared to $\mu_e$, as must to a lesser extent $\mu_5= \pi k_{\star}/\alpha$.  This in turn implies both that the growth rate $\Gamma_ {\rm CMI}$ would be very slow and that the total amount of electron energy available for conversion to magnetic field energy would be very small.  For example, for $k_\star \sim (\text{100 m})^{-1}$ one finds  $\mu_5\sim 10^{-6}\eV$ and $\Gamma_ {\rm CMI}\sim (\text{1 yr})^{-1}$.  

In order to find out what actually happens, we need to estimate how large $\mu_5$ gets in a core collapse supernova, and to do this we need to consider massive electrons.  Now the anomaly equation, \eq{anomaly}, is modified to include explicit chiral symmetry breaking due to the electron mass
\beq
\partial_\mu j^\mu_5= 2im\bar\psi \gamma_5\psi -\frac{\alpha}{2\pi} F_{\alpha\beta} \tilde F^{\alpha\beta}\ .
\eqn{anomalyb}
\eeq
It is not particularly simple to use this equation directly to compute rates in a plasma, since single particle asymptotic states are no longer eigenstates of chirality.
Instead it is useful to discuss electron helicity eigenstates, as helicity is exactly conserved for any electron mass in the absence of interactions.  For free massive electrons  in a multi-electron state of definite helicity $\ket{h}$,  the expectation value $\expect{h}{n_5}{h}$  is time-independent since $\ket{h}$ is a stationary state, despite $n_5$ not commuting with the Hamiltonian.  In fact, this expectation value is given by the sum of helicity times the magnitude of the  velocity ($|p|/E$) for each electron --- a result that goes smoothly to the $m=0$ limit, since in that limit all electrons have $|p|/E=1$ and helicity becomes synonymous with chirality. We can now turn on interactions and see how the evolution of $\ket{h}$ due to electron helicity flipping interactions leads to a time dependence of the expectation value of $n_5$, where
\beq
n_5(t) =  \int \frac{d^3 k}{(2\pi)^3} \left[f_+(k,t)-f_-(k,t)\right]\frac{|\bk|}{\omega_\bk} \ ,
\eqn{dn5}\eeq
 $f_\pm(k,t)$ being the electron occupation number in a state with momentum $k$ and $\pm$ helicity.   

 We make the assumption that deviations of $f_\pm(k,t)$ from  equilibrium are small, and use linear response with
\beq
f_\pm(k,t) = f(k) \pm \frac{\partial f(k)}{\partial \mu} \delta\mu_5(k,t)\simeq  f(k) \pm  \frac{\partial f(k)}{\partial \mu} \delta\bar \mu_5(t) \ ,
\eqn{lin}
\eeq
where $\delta\bar\mu_5(t)$ is $k$-independent and $f(k)$ is  the equilibrium Fermi-Dirac distribution
\beq
f(k) = \frac{1}{1+e^{-\beta (\omega_k - \mu_e)}}\ ,\qquad \omega_k = \sqrt{k^2 + m_e^2}\ .
\eeq
For the first part of \eq{lin} we simply assumed $|\delta\mu_5(k,t)|\ll \mu_e$ for all $k$, an approximation which will be seen to be self-consistent, as the equilibration of $n_5$ to zero due to electron helicity changing scattering is found to be much faster than the rate of change of $n_5$ arising from either the CMI or the weak interactions.  For the second part of \eq{lin} we used the fact that $\partial f/\partial \mu$ only has support for $|k-\mu_e|\lesssim T$, and assumed that $\delta\mu_5(k,t)$ was roughly independent of $k$ in this region, allowing the replacement $\delta\mu_5(k,t)\to \delta\bar\mu_5(t)$.  This latter assumption is justified by the fact that helicity preserving scattering will be fast (not suppressed by the small electron mass) and so the positive and negative helicity electrons will each be in independent approximate quasi-static thermal equilibrium. Given \eq{lin} we can express $n_5$ as
\beq
n_5(t)  \simeq  2\delta\bar\mu_5(t)\int \frac{d^3 k}{(2\pi)^3}  \frac{\partial f(k)}{\partial \mu} \frac{|\bk|}{\omega_\bk}\simeq \frac{\delta\bar\mu_5(t) \mu_e^2}{\pi^2} ,\qquad \Longrightarrow\qquad 
\frac{\dot n_5}{n_5} \simeq \frac{\delta{\dot{\bar{\mu}}_5}}{\delta\bar\mu_5}\ ,
\eqn{dmu5}\eeq
where again we made use of the fact that $\partial f(k)/\partial \mu$ is sharply peaked at $|k-\mu_e|\lesssim T$, with $m_e,T \ll\mu_e$.
We will refer to the contribution to $\dot n_5/n_5$ arising from electron helicity changing scattering as $-\Gamma_m$, since these contributions must vanish at zero electron mass. 

We find that helicity changing Rutherford scattering of electrons off the ambient protons to be the dominant contribution to $\Gamma_m$. Other contributions come from electron-electron scattering and  Compton scattering, but the former is expected to be suppressed relative to Rutherford scattering due to the fact that electrons are far more degenerate than protons,  while the latter is relatively suppressed since the proton density scales as $\mu_e^3$, while the ambient photon density scales as $T^3$, where $T$ is the temperature and  $T/\mu_e\lesssim  1/10$ during the core collapse.   From the Boltzman equation, in the approximation of \eq{lin}, we find
\beq
\frac{\partial_t \,\delta\mu_5(k,t)}{\delta\mu_5(k,t)} = -2  \frac{1}{2\omega_\bk}  \int \frac{d^3 k'}{(2\pi)^32\omega_{\bk'}}\,\frac{1+e^{\beta(\omega_\bk - \mu_e)}}{1+e^{\beta(\omega_{\bk'}- \mu_e)}}\,W(k,k')_{+-}
\eqn{dmudt}\eeq
where
\beq
W(k, k')_{hh'} = \int\frac{d^3 p\,d^3 p'}{(2\pi)^32 \omega_\bp(2\pi)^3 2\omega_{\bp'}}|\CM_{hh'}|^2(2\pi)^4\delta^4(p+k-p'-k') f(p')\left(1-f(p)\right) 
\eeq
for electron scattering with incoming and outgoing  momentum and helicity $(\bfk, h)$  and  $(\bfk',h')$ respectively, where $p$, $p'$ are the proton momenta, and $\CM_{hh'}$ is the Rutherford scattering amplitude averaged and summed  over incoming and outgoing proton spins. Neglecting proton recoil (suppressed by $\mu_e/M_p$) one finds
\beq
|\CM|^2_{+-} = 128 \pi^2 \alpha^2\frac{ E_\bp^2 m^2 \left(1-\cos\theta\right)}{\left(2 k^2(1-\cos \theta)+q_D^2\right)^2} 
\eeq
where $\theta$ is the scattering angle, $E_p$ is the proton energy,  and the inverse Debye screening length $q_D$ provides an infrared cutoff to the scattering process.  
Inserting this expression into \eq{dmudt} and evaluating at $k=\mu_e$, where $\partial f/\partial \mu$ is peaked, we find
\bea
\Gamma_m = -\left(\frac{\delta{\dot{\bar{\mu}}_5}}{\delta\bar\mu_5} \right)_\text{Ruth.}= -\left( \frac{\partial_t \,\delta\mu_5(k,t)}{\delta\mu_5(k,t)}\right)_\text{Ruth.}\Biggl\vert_{k=\mu_e}
 \simeq   \frac{\alpha^2  m_e^2 }{ 3\pi \mu_e} \left[\ln \frac{4}{x}-1\right]\ ,\qquad x\equiv \frac{q^2_D}{\mu^2_e}\ .
\eqn{gammam}\eea
Because proton degeneracy and recoil can be neglected, this result coincides with the simpler expression $\Gamma_m= n_p \sigma_{\rm R} (\mu_e)$, where $n_p$ is the proton density and $\sigma_{\rm R} (\mu_e)$ is the Rutherford cross section for electrons on the Fermi surface.   Noting that $q^2_D = 4\pi \alpha ~\partial^2 \Omega/\partial \mu_e^2$, where $\Omega$ is the total free energy of the plasma, and that at the fiducial density and temperature characteristic of the supernova, the electrons can be treated as degenerate and protons as non-degenerate, we find that $x = 4 \alpha( 1+ (\mu_e/3T))/\pi$. For $\mu_e=100\MeV$ and $T=30$ MeV we find $\Gamma_m \simeq 6 \times 10^{-8} \MeV \simeq 10^{14}/\text{ s}$.

The equation for the local evolution of the net helicity density including helicity flipping and electron capture rates is given by  
\beq 
\frac{1}{n_5}\frac{\partial n_5}{\partial t } = - \Gamma_B -  \Gamma_m + \frac{n_e}{n_5} \Gamma_w
\label{eq:n5t}
\eeq
where $\Gamma_w$ is the rate of depletion of the electron fraction $Y_e$ due to electron capture via charged current interactions during core collapse. Although $\Gamma_w$ is density and temperature dependent, and thus governed by complex supernova dynamics, a nearly model independent upper bound can be be derived by noting that the total change during core collapse $\delta Y_e \simeq 0.4$ occurs on a time scale that is greater than the free-fall timescale $t_{\text{free-fall}} \simeq 100$ ms. Therefore,  
\beq
\Gamma_w = \frac{\dot Y_e}{Y_e} <  10 ~\text{ s}^{-1} \,.
\eeq
However, simulations indicate that the typical value is $\Gamma_w \simeq 1\text{ s}^{-1}$ \cite{Burrows:2012} and we use this to make numerical estimates in the following calculations.  $\Gamma_m$ is the equilibration rate of $n_5$ due to  explicit chiral symmetry breaking by  the electron mass, given above in \eq{gammam}, and $\Gamma_B$ is the anomalous  depletion rate of $n_5$ due  the conversion of $n_5$ into magnetic field via the CMI.  We derived a formula for $\Gamma_B$ in the massless electron limit  in  \eq{hdot}, in the presence of a  chemical potential $\mu_5$.  In the realistic  case with nonzero electron mass, chirality is only approximately conserved,  and there is no chemical potential for chirality. Instead there is the effective  $\delta\bar\mu_5(t)$ computed in \eq{dmu5}. However, simply substituting this into  \eq{hdot} is  not valid in general, since the growing mode solution \eq{cmi} was derived assuming a constant $\mu_5$, which can be thought of as allowing the heat bath to provide an infinite source of energy for the growing magnetic field.  
 
The case where it is approximately valid to use  \eq{hdot}  with the substitution $\mu_5\to \delta\bar\mu_5(t)$ is when the CMI effect has a negligible effect on the background chiral density $n_5$.  We will investigate this regime and show that it is in fact a self-consistent solution during core collapse.  We first neglect $\Gamma_B$ in \eq{n5t}, in which case a fixed point solution is found where the slow production of $n_5$ from the weak interactions is balanced against the rapid equilibration of $n_5$ due to the nonzero electron mass:
 \beq
n_5=\frac{\Gamma_w}{\Gamma_m} n_e~ \sim 10^{-14} n_e\,.
\eqn{fp}\eeq   
Using \eq{dmu5}, this steady-state density corresponds to a very small time-independent effective chemical potential 
\beq
\delta\bar\mu_5 = \frac{\pi^2 n_5}{\mu_e^2} =  \frac{\pi^2 n_e\Gamma_w}{\mu_e^2 \Gamma_m}\simeq \frac{ \mu_e}{3} \frac{\Gamma_w}{ \Gamma_m} \sim \frac{1}{3}10^{-14} \mu_e
\eqn{mu5val}\eeq
We can now use this steady state value to compute the rate of magnetic field production, $\Gamma_B$, as well as the length scale of the unstable mode, $k_\star$. We find that  $k_\star^{-1} = \pi/(\alpha\delta\bar\mu_5 )\sim 250 \text{ m}$ for $\mu_e \simeq 100$ MeV which is astrophysically interesting.  Using the above expression for $\delta\bar\mu_5 $ in \eq{hdot} and \eq{cmi} to compute $\Gamma_B$, we find
\beq
\Gamma_B(t) =\frac{2\alpha^2}{\pi^2\sigma} \frac{ \delta\bar\mu_5}{n_5}  B_\star(t)^2 =\frac{2\alpha^2 }{\sigma \mu_e^2 }  B_\star(t)^2 =\frac{2\alpha^2 }{\sigma \mu_e^2 } B_\star(0)^2  e^{ 2 t\Gamma_\text{CMI}}\ ,\qquad \Gamma_{\rm CMI} =\frac{\alpha^2 \delta\bar\mu_5^2}{\pi^2\sigma}\ .
\eeq

 The above expression for $\Gamma_B$ is only valid to the extent that  $\Gamma_B\ll \Gamma_w$, or else the fixed point solution \eq{fp}  -- obtained by ignoring the effects of $\Gamma_B$ -- will not be correct.   Such an inequality will  break down eventually due to the exponential growth of $B_\star(t)$ if the  prefactor proportional to the seed field $B_\star(0)^2$ at wavenumber $k_\star$ is sufficiently large and the time scale $\Gamma^{-1}_{\rm CMI} $ is sufficiently short compared to the duration of core collapse and the concomitant electron capture.  
 
Both the prefactor and the exponential growth rate depend on  the electrical conductivity $\sigma$, which  is quite high in the supernova plasma.   By assuming that protons are non-degenerate and uncorrelated we derive a lower bound 
\beq 
\sigma\gtrsim \sigma_\text{min} = \frac{\mu_e}{4 \alpha} \left[\ln \frac{4}{x}-1\right]^{-1}\,,\qquad x\equiv \frac{q^2_D}{\mu^2_e}\ .
\eeq
which implies that
\beq 
\Gamma_B(0) =\frac{2\alpha^2 }{\sigma  \mu_e^2 }  B_\star(0)^2 \lesssim  \frac{8 \alpha^3}{\mu_e^3}  \left[\ln \frac{4}{x}-1\right]  B_\star(0)^2  \sim \Gamma_m \times\left(\frac{B_\star(0)}{5\times 10^{14}\text{ G}}\right)^2 \, 
\eeq
and
\beq
\Gamma_{\rm CMI} =\frac{\alpha^2 \delta\bar\mu_5^2}{\pi^2\sigma}\lesssim \frac{4\alpha^3}{9\pi^2}   \left[\ln \frac{4}{x}-1\right] \left(\frac{\Gamma_w}{\Gamma_m}\right)^2  \mu_e\sim 6\times 10^{-34}\MeV\sim 10^{-12}~\text{s}^{-1} \ .
\eeq
We see that at the beginning of the collapse, our assumption that $\Gamma_B$ may be neglected compared to $\Gamma_m$ is justified unless the initial seed field $B_\star(0)$ is already very large, around $10^{15}$~G.  Furthermore, for more moderate initial magnetic fields, the extremely slow growth rate means that no exponential enhancement of the magnetic field occurs during the few seconds of core collapse.  Finally it should be noted that if $\Gamma_B$ was ever large compared to $\Gamma_m$, that would only serve to drive $n_5$ smaller, slowing the process down and driving it to smaller wave number $k_\star$.  It is remarkable that the relatively large value obtained for $\Gamma_m$ -- which is proportional to $m_e^2$ -- is responsible for damping out the chiral magnetic instability. To our knowledge, this is the first time that the fact that the electron is not massless has been shown to play a critical role in the structure and evolution of neutron stars.

In closing we comment on the idea that a permanent instability could persist in cold neutron matter due to the the neutral current interaction between electrons and neutrons, proportional to $G_F (\bar e \gamma^\mu\gamma_5 e)(\bar n\gamma_\mu n)$.  It has been observed that in mean field theory, this term gives an effective contribution to the electron dispersion relation that resembles a chiral chemical potential, $(G_F n)(\bar e \gamma^0\gamma_5 e)$,  where $n$ is the neutron density.  That such a term could lead to a magnetic instability  was proposed in ref.~\cite{Boyarsky:2012ex}, and considered but discarded much earlier by Vilenkin \cite{Vilenkin:1980ft}, who also considered the effects of rotation.  While an attractive idea for generating the large magnetic fields observed in magnetars, we note the absence of an energy source for the growing magnetic field in this scenario, making Vilenkin's conclusion that such a mechanism does not work more intuitively plausible.  This is to be contrasted with the scenario considered in this paper, where the growth of the helical magnetic field is powered by gravitational energy released during core collapse and temporarily stored in fermi energy of the left handed and right hand electrons, which are temporarily out of thermal equilibrium with each other; a mechanism that fails because the electron mass never allows them to depart very far from equilibrium.  Apparently what is needed to explain magnetars is a more efficient mechanism for transferring the gravitational energy released during collapse into electromagnetic energy.

\begin{acknowledgments}
We would like to thank Edward Witten for useful discussions, and Jose Pons for alerting us to recent work on the chiral magnetic instability in neutron stars.  The work of D. G., D. K. and S. R. was supported by the DOE Grant No. DE-FG02-00ER41132. The work of S. R. was also supported by DOE Topical Collaboration to study neutrinos and nucleosynthesis in hot dense matter. This material is based upon work supported by the National Science Foundation Graduate Research Fellowship under Grant No. DGE-1256082.
\end{acknowledgments}

\bibliographystyle{apsrev4-1}
%\bibliography{cme}
%%%%%%%%%%%%%%%%%%%%%%
%
\end{document}